\documentclass[10pt,aps,pra,twocolumn,showpacs,superscriptaddress,floatfix]{revtex4-1}  
\usepackage{graphicx}  
\usepackage{dcolumn}   
\usepackage{bm}        
\usepackage{amsmath}
\usepackage{amssymb}   

\usepackage[utf8]{inputenc}
\usepackage{xfrac}

\newcommand{\create}[2]{\hat{#1}^{\dagger}_{#2}}
\newcommand{\annihilate}[2]{\hat{#1}^{\phantom{\dagger}}_{#2}}
\newcommand{\n}[1]{\hat{n}_{#1}}
\newcommand{\id}{\mathrm{d}}

\begin{document}

\title{Using superlattice potentials to probe long-range magnetic correlations in optical lattices}
\author{Kim G.~L.~Pedersen}
\affiliation{Institut f\"{u}r Theorie der Statistischen Physik, RWTH Aachen University}
\author{Brian M.~Andersen}
\affiliation{Niels Bohr Institute, University of Copenhagen}
\author{Georg M.~Bruun}
\affiliation{Institute of Physics and Astronomy, Aarhus University}
\author{Anders S.~S{\o}rensen}
\affiliation{Niels Bohr Institute, University of Copenhagen}
\email{anders.sorensen@nbi.ku.dk}

\date{\today}

\begin{abstract}
In \citet{Pedersen2011} we proposed a method to utilize  a temporally dependent superlattice potential to mediate spin-selective transport, and thereby probe long and short range magnetic correlations in optical lattices.
Specifically this can be used  for detecting antiferromagnetic ordering in repulsive fermionic optical lattice systems, but more generally it can serve as a means of directly probing correlations among the atoms by measuring the mean value of an observable, the number of double occupied sites.   Here, we provide a detailed investigation of the physical processes which limit the effectiveness of this ``conveyer belt method''. Furthermore we propose a simple ways to improve the procedure, resulting in an essentially perfect (error-free) probing of the magnetic correlations. These results shows that suitably constructed superlattices constitute a promising way of manipulating atoms of different spin species as well as probing their interactions.
\end{abstract}

\maketitle

\section{Introduction}
Cold atoms in optical lattices are  powerful ``quantum simulators'' for  strongly correlated many-body systems in periodic potentials. The experimental study of these systems has resulted in a range of breakthrough  results in the last two decades, including the observation of the Mott-superfluid transition~\cite{Greiner2002,Jordens2008,Schneider2008}, fermionic pairing~\cite{Chin2006}, Dirac points~\cite{Tarruell2012}, and various topological band structures~\cite{Atala2013,Aidelsburger2013,Miyake2013}. 
Fermionic atoms in optical lattices can be  accurately described by the single Fermi Hubbard model~\cite{Jaksch1998,Bloch2008}, and a major goal of current research in this area is to explore this model at low entropies where it, e.g., is expected to exhibit antiferromagnetic order. Moreover, there is strong numerical evidence that the Fermi Hubbard model captures the essential physics leading to high-$T_c$ superconductivity~\cite{Zanchi1996, Halboth2000, Honerkamp2001, Maier2005, Aimi2007, Romer2015}.
Using Bragg spectroscopy~\cite{Corcovilos2010}, one recently observed the onset of antiferromagnetic correlations reaching temperatures as low as $1.4$ times the critical temperature for antiferromagnetic ordering~\cite{Hart2015}, and such correlations were also observed along one dimension in an anisotropic lattice~\cite{Greif2013}.  Several alternative  methods  have been proposed for observing magnetic correlations, including time-of-flight~\cite{Altman2004,Bruun2008,Andersen2007}, and light-polarization experiments~\cite{Eckert2008,Bruun2009}.
A common problem for most of these probes~\cite{Greif2013,Altman2004,Bruun2008,Andersen2007,Eckert2008} is that they are only indirectly probing, e.g., the antiferromagnetic ordering. Since they inherently rely on dynamics in momentum space~\cite{Corcovilos2010,Altman2004,Bruun2008,Andersen2007} or on averaging over large areas~\cite{Bruun2009}, they are not very sensitive to direct spatial correlations of the spins such as the antiferromagnetic ordering of neighboring atoms.   

In \cite{Pedersen2011}, we demonstrated how a moving superlattice potential can be used like an atomic conveyor belt to selectively transport one atomic spin component a desired number of lattice sites. In this way, one directly can measure the spatial dependence of the spin-spin correlation function, and in particular  detect both the long range order as well as short range correlations which may mark the onset of long range correlations. Importantly, in this proposal the desired correlation function is directly mapped into an observable quantity, the number of double occupied sites. This means that there is no need to average over several shots of the experiment since correlations can be directly measured in a single shot. This will be a particularly important property if the details of the experiment is fluctuating from shot to shot so that not every instant of the experiment show the desired ordering.
In this article we investigate the scheme of Ref.~\cite{Pedersen2011} in further detail, improving and optimizing the different steps. 

A detailed description of the superlattice conveyer belt (SCB) probing scheme of Ref.~\cite{Pedersen2011} is presented in Sec.~\ref{sub:IIA}.  An essential step in the procedure is the velocity  with which the superlattice is moved through certain avoided crossings to ensure the correct adiabatic or diabatic behavior in order to obtain the desired selective hopping of atoms. 
We analyze this in Sec.~\ref{sub:IIB} by mapping the system to a Landau-Zener tunneling between nearly degenerate levels on neighboring sites. Based on this mapping we identify suitable parameter regimes where the procedure can have a good performance. 
Then in Sec.~\ref{sub:IIC}  we consider unwanted hopping between lattice sites. Again we reduce this to a simple model by identifying it with decoherence caused by the width of the  Bloch bands. Based on the simple model we find  suitable regimes of  superlattice amplitudes where we predict the scheme to have a good performance. In Sec.~\ref{sub:IID} we consider  another source of erroneous  hopping, due the presence of  resonances of higher Bloch bands as the superlattice is moved. These hoppings are a major source of error for the original procedure of  Ref.~\cite{Pedersen2011}, but we show that they can be almost eliminated  by  ramping down the superlattice amplitude in between the desired hoppings. 
Finally, in Sec.~\ref{sec:III} we investigate the validity of our conclusions from the simplified models with  a numerical calculation. These simulations show that the modified procedure can measure the spatial dependence of the spin-spin correlation function essentially without error both for short and long distances. 

\section{The superlattice conveyer belt} \label{sec:II}
\subsection{Ideal procedure} \label{sub:IIA}
In the following, we assume that the dynamics in the transverse $y$ and $z$ directions is frozen by a tight transverse lattice potential, so that the problem is effectively one-dimensional involving only the $x$-direction. 
We consider atoms trapped in an optical lattice described by a potential of the form 
\begin{align}
V_A(x) = A \sin^2 (2 \pi x / \lambda),
\end{align} 
where $\lambda$ is the wavelength and $A$ denotes the strength of the potential. With this potential it is convenient to introduce the lattice recoil energy, $E_R = h^2 / 2 m \lambda^2$ as a natural energy scale while time is measured in units of $t_R = h/E_R$~\cite{Jaksch1998}. As a specific example, we consider atoms in two different spin (hyperfine) states, but the method may be generalized to atoms with more spin states.   At time $t=0$ the atoms are prepared  in the many-body state we wish to probe. How we arrive at this is not important for the analysis below and  it may, e.g., be an equilibrium state or the result of a dynamical evolution of a strongly interacting system. From the time we apply the SCB probe, however,  we assume that the interactions among the atoms are sufficiently weak that we can ignore them. 
This can be the case if the interactions are sufficiently weak that they can be ignored on the time scale of the SCB, which can be much faster than the preceding dynamics, or if the interactions are turned off before the SCB using for instance a Feshbach resonance~\cite{Chin2010}. 
 
The first step of the probing procedure is a spin-selective Raman pulse which excites one of the spin states to 
 the first excited vibrational level, while leaving the other spin in the vibrational ground state~\cite{Bouchoule1999}. A time-dependent superlattice potential with twice the wave length of the form 
\begin{equation}
 V_B(x,t) = B(t) \sin^2 \left\{\pi [x - x_B(t)]/ \lambda \right\}
 \label{SuperlatticeEq}
\end{equation} 
is then applied. Here $B$ denotes the strength of the superlattice potential and $x_B(t)$ is a time dependent displacement of the superlattice potential, which we use to create the conveyor belt below. This potential needs to be turned on sufficiently fast that atoms do not have time to tunnel to neighboring sites, but at the same time sufficiently slowly that they adiabatically follow the vibrational levels at each site. 
  
\begin{figure}[tb]
  \begin{center}
    \includegraphics[width=\columnwidth]{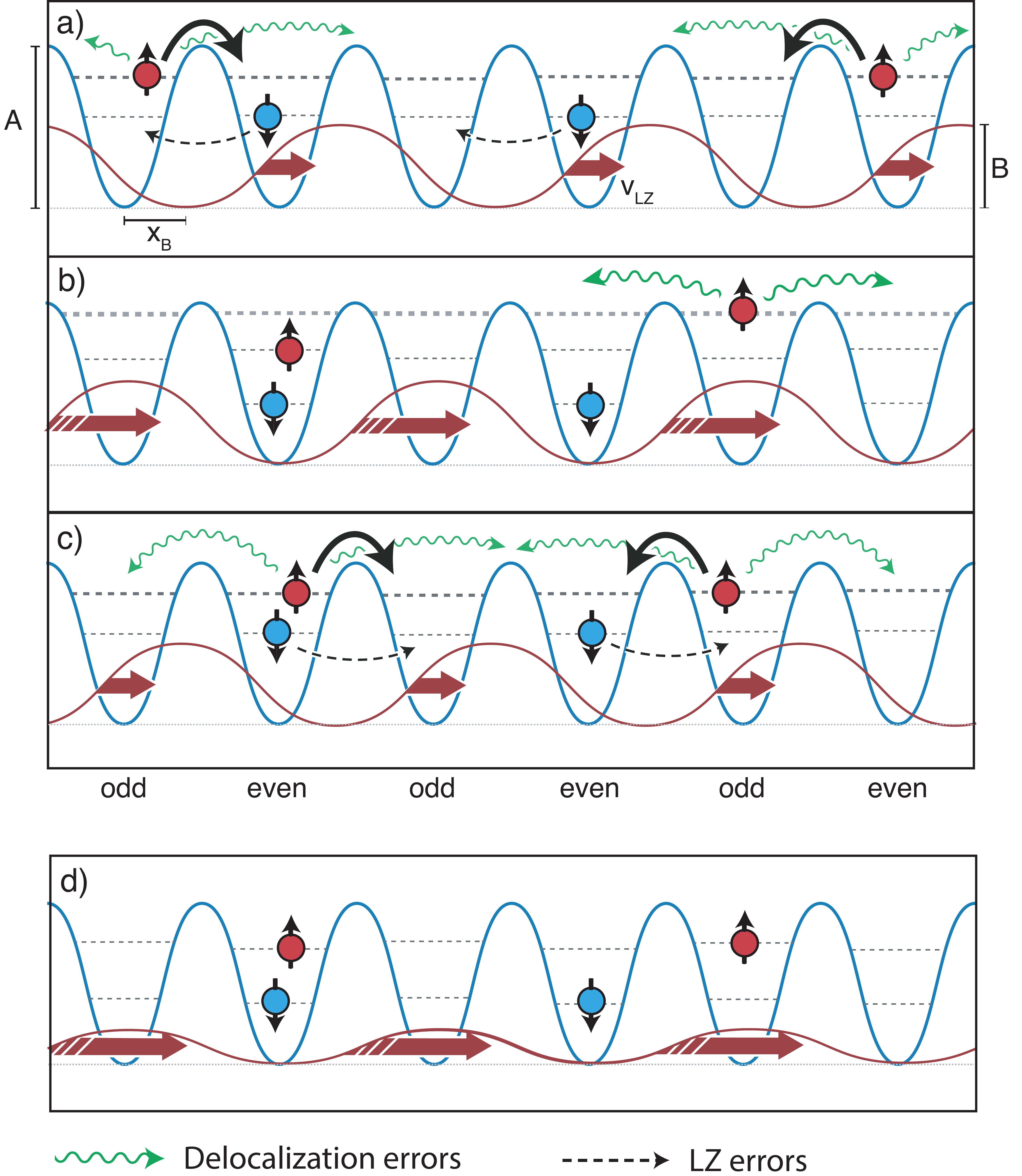}
  \end{center}
  \caption{Snapshots of the superlattice ``conveyer belt'' procedure. (a) Superlattice potential $V_B$ (red line) placed near the degeneracy point where neighboring wells are  degenerate. The combined potential creates distinct double well structures, e.g., counting from the left, minima number one and two of $V_A$ (blue line) are connected, whereas the tunneling from minima two to three is suppressed by $V_B$. Here the superlattice velocity is tuned such that atoms in the first vibrationally excited state (red circles) cross adiabatically moving along with the conveyor belt (thick arrows), whereas atoms in the ground state (blue circles) remain stationary. Atoms in the odd sub-lattice move in the opposite direction of atoms starting in the even sub-lattice. The dashed arrows show potential Landau-Zener errors, while wiggly arrows show delocalization errors due to the presence of multiple double wells. (b) After the crossing, the superlattice is moved quickly to the next degeneracy point. Note that atoms starting in the even sub-lattice are only weakly confined allowing them to delocalize easily as indicated by the stronger wiggly lines. (c) The next  degeneracy point, which transports atoms in the excited states to the next neighboring site. Here a double well system is formed by, e.g., minima number two and three of $V_A$. (d) Modified superlattice procedure with a decreased superlattice strength during the displacement.
  In the modified SCB this step replaces part (b).}
   \label{fig:procedure}
\end{figure}

\begin{figure}[tb]
  \begin{center}
    \includegraphics[width=\columnwidth]{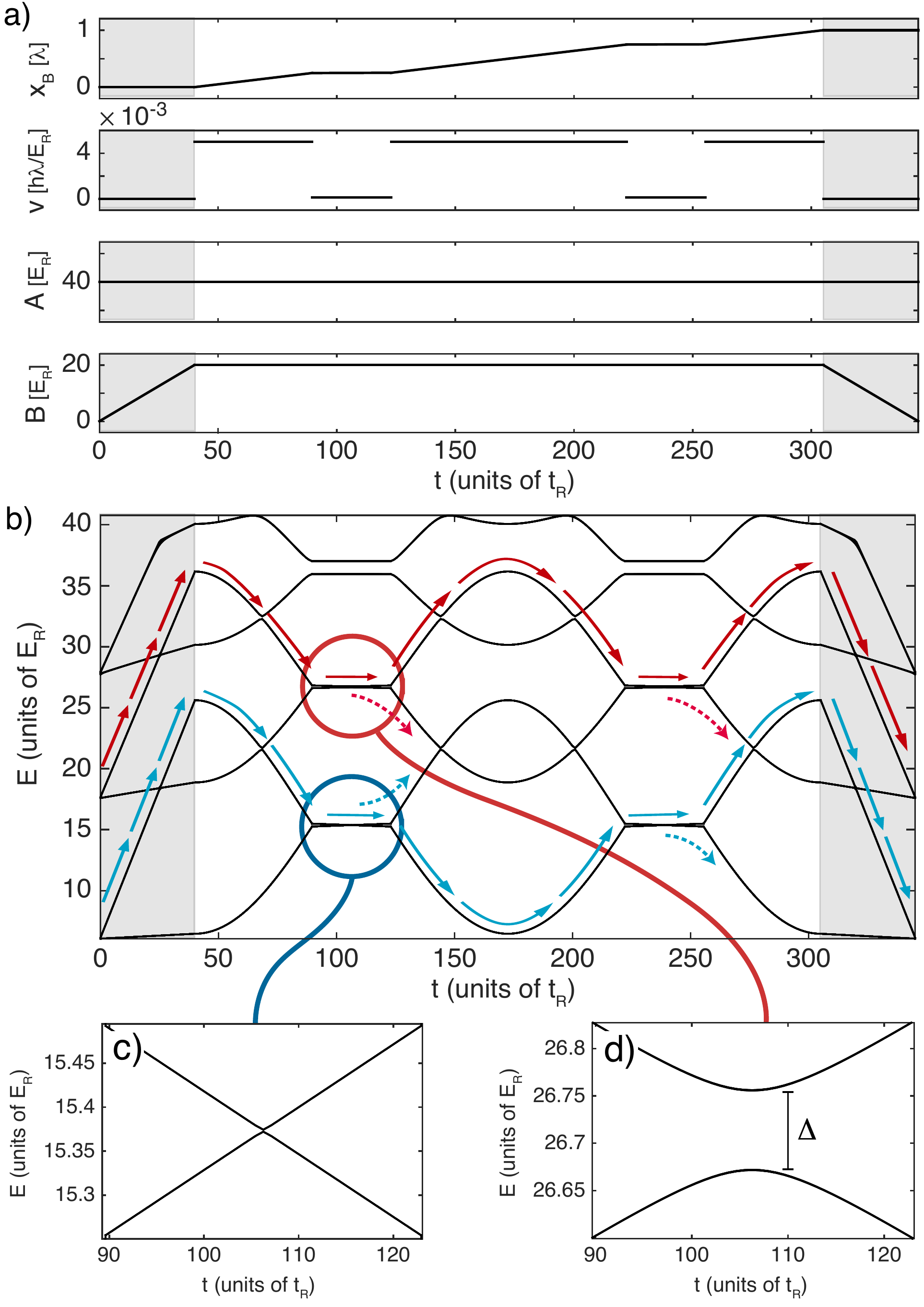}
  \end{center}
\caption{The original superlattice procedure with $A = 40 E_R$  and $B = 20 E_R$ proposed in Ref.~\cite{Pedersen2011}. (a) Superlattice-lattice parameters as a function of time, when the superlattice conveyor belt moves atoms two sites away from their starting point. Note that $v$ is small but non-zero in the intervals around $t=100 t_R$ and $t=250 t_R$, whereas it is exactly zero around $t=0$ and $t=330 t_R$ (b) The instantaneous Bloch eigenenergies for all Bloch wavevectors $k$ as a function of time. The Bloch eigenenergies for each wavevector almost coincide, so that the Bloch bands look like lines. The dashed arrows indicate the possible Landau-Zener errors. Zoom of the eigen-energies  near the avoided crossing for (c) the ground state and (d) the first excited state. The superlattice velocity is chosen such that the ground states cross diabatically, while the first excited states cross adiabatically.}
\label{fig:SCB-A40B20}
\end{figure}

The key idea in our probing procedure is to make the time-dependent potential act as a ``conveyor belt'' for  atoms populating the first excited vibrational level. To do this we use the total potential $V_A+V_B(t)$ to  create an array of almost isolated double well systems. This happens at $x_B\approx\lambda/4$, where the maxima/minima of the superlattice potential $V_B(t)$ are close to the maxima of the original lattice $V_A$, as illustrated in Fig.~\ref{fig:procedure} (a).  We shall henceforth refer to this position as the degeneracy point. At this position atoms can not easily tunnel between lattice sites of $V_A$, which have a maximum of $V_B$ in between them. On the contrary tunneling is not suppressed between sites which have a minimum of $V_B$ in between them.  Thus effectively the lattice potentials creates an array of coupled double wells. In these double wells the atoms in the vibrationally excited states move between  degenerate  levels located in the two wells, while the atoms in the vibrational ground state will remain almost stationary due to the higher tunneling barrier. To turn this into a conveyor belt we need to interchange the populations of the first excited states of the two wells. We do this by moving the superlattice potential across the degeneracy point. In  Fig.~\ref{fig:procedure} (a) the superlattice $V_B$ is displaced slightly to the left of the degeneracy point. This means that the right well is slightly higher in energy. Moving the superlattice to the right exchanges the roles so that the left well is now higher in energy. If the superlattice is moved slowly enough the populations will follow \emph{adiabatically} and thus be interchanged.
Due to the difference in the tunneling strength one can tune the superlattice velocity  $v_B = \id x_B/ \id t$ such that only the atoms in the first excited vibrational level follow adiabatically, whereas atoms in the ground state will remain stationary as illustrated in Fig.~\ref{fig:procedure}~(a).  This corresponds to the atoms in the first excited states following the instantaneous Bloch level spectrum adiabatically, as opposed to atoms in the ground state, which cross the level spectrum \emph{diabatically}. This is illustrated in Fig.~\ref{fig:SCB-A40B20} around $t=100 t_R$ (see below for how we obtain this plot). 
 
The above procedure moves atoms in the first excited vibrational level by a single position. To make the conveyor belt move the atoms over larger distances, we quickly displace [Fig.~\ref{fig:procedure}~(b)] the superlattice potential $V_B$ to the next degeneracy point where a new array of double wells are created, but now with each well coupled in the opposite direction  [Fig.~\ref{fig:procedure}~(c)].   
Here it is essential that the superlattice is moved quickly in-between the degeneracy points, since we need to ensure that  atoms remain at the same positions during this period. In a Bloch  band picture this corresponds to the diabatic level crossings illustrated around e.g. $t\approx150 t_R$  and $t\approx200 E_R$ in Fig.~\ref{fig:SCB-A40B20} and is discussed in detail in Sec.~\ref{sub:IID}.
Repeated crossings of degeneracy points then move the atoms in the first excited state down the lattice. The number of crossings then directly translates into the number of translations of the atoms.
Note that atoms on odd (even) sites move to the left (right) during this procedure. 
 
At the end of the procedure, the superlattice is turned off and the number of doubly occupied sites is measured. This can for instance be done by transferring atoms from the excited vibrational level to the ground state and merging the doubly occupied states into molecules through a Feshbach resonance or photo association~\cite{Greiner2002,Jordens2008,Schneider2008,Greif2011} followed by a measurement of the number of molecules. The fraction of molecules to atoms then directly measures the long-range spin-correlations in the system as discussed in Ref.~\cite{Pedersen2011}.  Specifically in the ideal limit the molecular fraction, $P_{mol}(m)$, after $m$ applications of the conveyor belt, can be related to the spin correlation function through,
\begin{align}
	P_{mol} &= 1/4 - \langle s^z_{i+m} s^z_{i} \rangle.
\end{align}
Here $\langle s^z_{i+m} s^z_{i} \rangle$ is the spin-spin correlation function between spins, located $m$ sites apart. 

In total the SCB is created by a  number of different sequences with different parameters. The time evolution of the various parameters entering into the evolution is shown in Fig.~\ref{fig:SCB-A40B20} (a) for the particular example analyzed in Ref.~\cite{Pedersen2011}.
In the following, we analyze in detail the different steps of the SCB procedure in order to minimize the main sources of error. We furthermore 
improve the method by allowing the amplitude $B$ of the superlattice potential to vary with time. We show that this may  significantly reduce the error of the protocol so that it can operate with almost no error.

\subsection{Optimizing the superlattice velocity} \label{sub:IIB}
The crucial step in the SCB procedure involves the neighbor well degeneracy shown in Fig.~\ref{fig:procedure}(a) and (c) and Fig.~\ref{fig:SCB-A40B20}, where all atoms in the first excited vibrational state  move to their neighboring site, while all atoms in the ground state  remain at their initial site. 
If however, the superlattice moves too slowly close to the degeneracy point, atoms in the ground state will also tunnel to their neighbor sites. If, on the other hand, the superlattice is moved too fast, atoms in the first excited state do not have sufficient time to tunnel to the neighboring site. It follows that there is an optimum superlattice velocity, where the degeneracy point is traversed neither too slow nor too fast, in order to ensure that most of the atoms in the first excited state move to the neighbor site, while most atoms in the ground state stay put.

We will  now use Landau-Zener theory for adiabatic passage in two level systems to estimate the optimum velocity. Our system, however, does not consists of two level systems  and we need to extract effective parameters to use in the Landau-Zener theory from the description of the full system.
Regardless of the position of the superlattice, the instantaneous potential is always periodic with a period $\lambda$ set by the periodicity of the superlattice. This means that we can use  
Bloch's theorem which specifies that the eigenstates with  energy $E_k^l$ of a one-dimensional periodic system are Bloch waves, ${\langle x|\Phi_k^l \rangle= e^{i k x} u^l_k (x)}$, parametrized by their  wave number, $k$, and their vibrational level, $l$. 
We therefore calculate the Bloch level spectrum numerically for various lattice-superlattice amplitudes $(A,B)$ and displacements $x_B(t)$. 
This is done by numerical diagonalization of the Hamiltonian discretised on a set of 256 real space positions within the unit cell.
The result of this calculation is shown in Fig.~\ref{fig:SCB-A40B20} (plotted as a function of time rather than displacement). Note that while the figure seems to consist of discrete lines, each line is in fact a whole band of energies corresponding the various $k$-values, but the width of each band is too narrow to be visible. 

Close to the degeneracy point, the plots in  Fig.~\ref{fig:SCB-A40B20} (c) and (d) show that the system is well described by two level systems corresponding to tunneling between one vibrational level from each well. Furthermore they have nearly linear asymptotics far away from the crossing, i.e. ${E_2 - E_1 = \beta x}$ as required by the Landau-Zener description.
At the avoided crossing, we approximate the system by two states, of energies $E_1$ and $E_2$, tunnel coupled by a matrix element set by the gap $\Delta$ between the levels at the degeneracy point.  We extract the energy gap $\Delta$ and the slope $\beta$ from the energy spectrum for each Bloch wave-vector.
The probability of an atom moving \emph{diabatically} from one instantaneous eigenstate to the other through the avoided crossing is then given by the Landau-Zener transition amplitude~\cite{Zener1932,Landau1932}
\begin{align} \label{eqn:plz}
	P_{LZ} = e^{-\pi \Delta^2/2 \alpha \hbar }.
\end{align}
Near the  degeneracy point, the level spacing is proportional to the displacement of the superlattice and we thus have  ${\alpha = v_B \beta}$. A perfectly working SCB will have the ground state eigen-energies crossing diabatically, ${P_{LZ}(gs) \approx 1}$, while the first excited state energies cross adiabatically, ${P_{LZ}(ex) \approx 0}$. 
The transition probability can then be calculated from Eq.~\eqref{eqn:plz} and the total success probability when crossing the degeneracy point is defined as the average over all the Bloch wave vector
\begin{equation}
	P_{\text{crossing}} = \frac{1}{N} \sum_{k} [P_{LZ}(gs) - P_{LZ}(ex)].
	\label{eqn:PNWD}
\end{equation}
From this the optimum superlattice velocity is found by maximizing $P_{\text{crossing}}$. 
  
\begin{figure}[tbh]
  \begin{center}
    \includegraphics[width=\columnwidth]{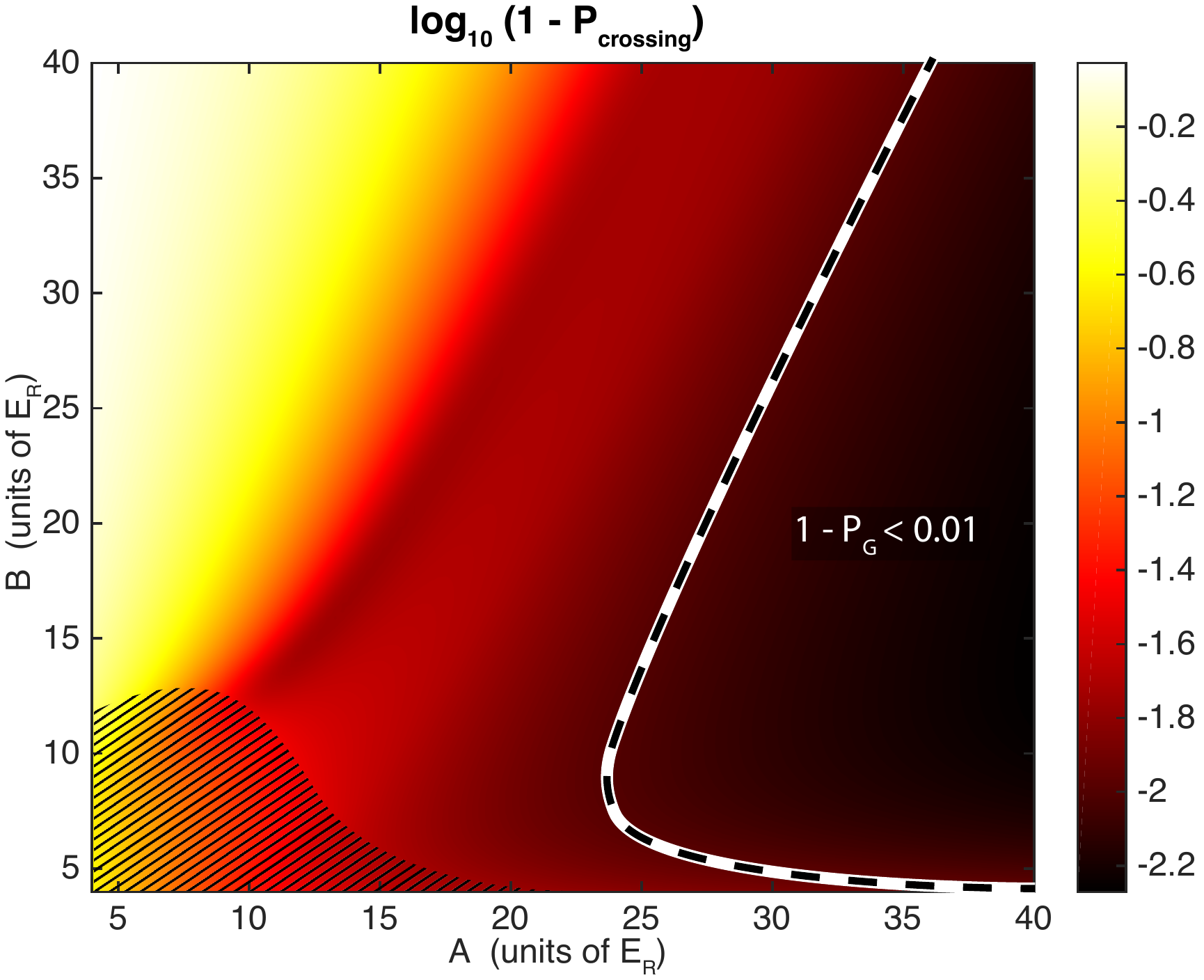}
  \end{center}
  \caption{Logarithmic color plot of the error probability of the Landau-Zener procedure as obtained from Eqs.~\eqref{eqn:plz}-\eqref{eqn:PNWD}. 
  The white dashed line encloses the  region with an error less than one percent. The hatched area marks parameters where the dynamics cannot be described by Landau-Zener physics.}
  \label{fig:plz}
\end{figure}

In Fig.\ \ref{fig:plz}, we plot the resulting optimal error probability, $1-P_{\text{crossing}}$ for lattice amplitudes  $(A, B)$ in the range ${[4 E_R, 40 E_R]}$.
Note that this estimate requires that the Landau-Zener description is a sufficiently good approximation of the dynamics. We therefore have to exclude an area in the lower left corner defined heuristically by $\Delta > B/4$, where we find that the curves do not have near linear asymptotics and is thus not well described by the Landau-Zener formula. This is, however, anyway not the regime we are interested in, since the scheme does not have a good performance for these parameters. On the other hand in the regime where the performance is good, we find the Landau-Zener approximation to be justified. From  the figure we identify a region in $(A,B)$ space where the error is less than one percent at the optimal velocity. This requires a large $A$, and $B$ has to be larger than a minimal value which is nearly independent of $A$ but smaller than an upper limit set by $A$.

Below we find a more stringent requirement for the lower limit for $B$ and we shall therefore not discuss it further. The requirements on the value of $A$ has a simple physical explanation: A large $A$ is required in order to have a large energy separation between the ground state and the first excited state, and therefore a large difference in the tunneling rates of the two levels.  On the other hand  a strong potential $V_B$ pushes the minima of the double well potential closer to the barrier in the middle and thus effectively lowers the tunneling barrier. In fact for $B\geq 4A$ the bump in the middle of the double well disappears. Hence in order for the system to be meaningfully described as  a double well, there is an upper limit on $B$ which grows with increasing $A$.

\subsection{Bloch level delocalization close to the neighbor well degeneracy} \label{sub:IIC}
The SCB procedure is based on the fact that at the degeneracy points, the lattice can be approximated by an array of isolated double wells. 
This approximation is, however, not exact and  atoms can still hop between different double wells which leads to delocalization errors as indicated by the green arrows in Fig.~\ref{fig:procedure}. 
This delocalization is most significant for the first excited vibrational mode, and can be used to construct a simpler but also less controlled probe of ordering in optical lattices~\cite{Pedersen2012}. For the SCB probe, on the other hand, the delocalization is a source of error and in this section, we shall investigate the effect of it. 

In order to ease our analysis, we label the states of the $V_A$ lattice by a superlattice site index $n$ and a compound index $l = (m, \eta)$, with $m$ referring the vibrational level and $\eta = e/o$ to the even or odd sub-lattice. Note that this notion of Wannier states being localized in the original lattice $V_A$ is strictly speaking only meaningful away from the degeneracy point, but this suffices for our present purpose.

The Wannier wave functions localized around the $n$'th superlattice site can be constructed from the instantaneous Bloch waves,
\begin{align}
	\Psi_n^l (x, t_0) = \frac{1}{\sqrt{N}} \sum_k e^{- i k n \lambda} \Phi^l_k (x, t_0).
\end{align}
Loosely speaking, the delocalization error is due to the the spread in energy of the Bloch band which means that the different wave numbers have different phase evolution. This difference in phase evolution is just the mathematical description of the atoms tunneling to different lattice sites. While the width of the Bloch band may seem negligible near the degeneracy point in Fig.~\ref{fig:SCB-A40B20}, one must remember that it should be compared to the slow pace set by the optimum Landau-Zener velocity.

We define the localization probability as the overlap between the initial first excited state $\Psi_n^{(1, \eta)} (t_0)$, and the adiabatically time-evolved state, $\Psi_n^{(1, \bar{\eta})} (t_1)$, which is transferred from the neighbor well, $\bar{\eta}$, during the crossing of the degeneracy point. We will show that this amounts to an integral over the energies within the Bloch band,
\begin{align}
 	P_{\text{local}} &= \sum_\eta \Big|\langle \Psi_n^{(1,\eta)} (t_0) | \Psi_n^{(1,\bar{\eta})} (t_1) \rangle\Big|^2 \label{eqn:Plocal} \\ &= \sum_\eta \Big|\frac{1}{N} \sum_k e^{-i \int_{t_0}^{t_1} E^{(1,\eta)}_k(t') \id t'}\Big|^2.
 	\label{eqn:Plocal2}
\end{align} 
To see this, we first write down the adiabatically time-evolved state, 
\begin{align}
	\Psi_n^l (x,t_1) = \frac{1}{\sqrt{N}} \sum_k e^{- i k n \lambda} e^{-i \int_{t_0}^{t_1} E_k^l (t') \id t'} \Phi^l_k (x, t_1).
\end{align}
Due to the spatial symmetry of the initial and final state the instantaneous Bloch eigenstates are related in the following way, ${\Phi^{(m,\eta)}_k (x,t_0) = \Phi^{(m,\bar{\eta})}_k (x, t_1)}$. Combining those two facts allow us to turn Eq.~(\ref{eqn:Plocal}) into Eq.~(\ref{eqn:Plocal2}) with only a bit of algebra.

From the localization probability, we define the localization error, $1-P_{\text{local}}$. As seen from Eq.~(\ref{eqn:Plocal2}) it depends on the time available for particles to tunnel to a different site. To extract this time,  we need to consider how wide a region around the degeneracy point we need to traverse. The superlattice velocity is slowed down in a range of superlattice positions $x_B \in [\lambda/4 - w/2, \lambda/4 + w/2]$, where suitable widths $w$ are determined heuristically: Assuming that the potential minima do not change significantly from the position of the minima of the potential $V_A$ the derivative of the energy is $\beta\sim 2\pi B/\lambda$. Hence by setting  $w = 2 \Delta \lambda / B$ the energy separation at the endpoints is $E_2-E_1\sim2\pi\Delta$ which is much larger than the splitting $\Delta$ at the avoided crossing. Combined with the optimum superlattice velocity 
obtained in the previous section, we use this to calculate the transversal time $t_1-t_0$ and then the
delocalization error for a variety of the lattice-superlattice amplitudes $(A,B)$. 

\begin{figure}[tb]
  \begin{center}
    \includegraphics[width=\columnwidth]{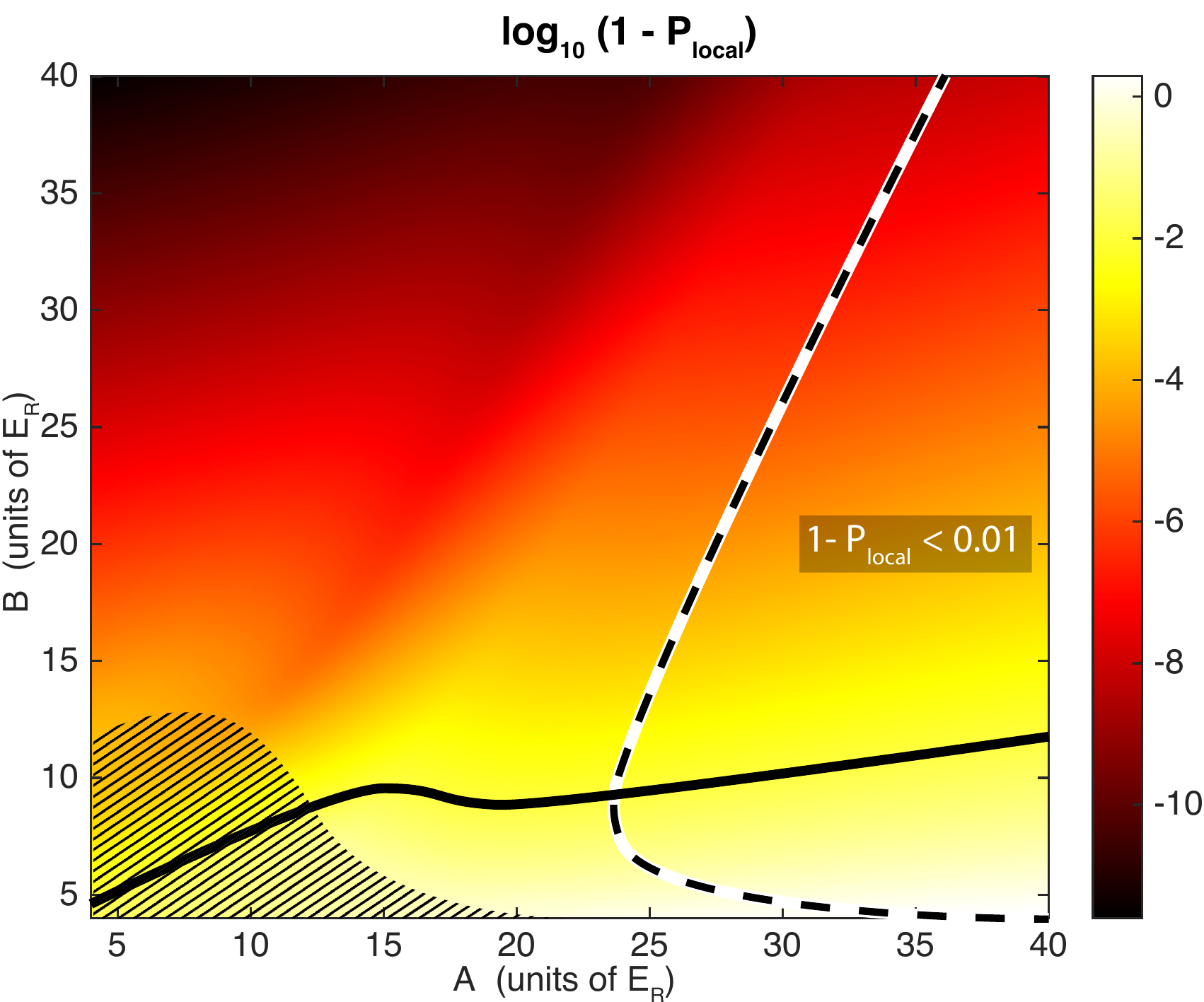}
  \end{center}
  \caption{The delocalization -- as defined by Eq. \eqref{eqn:Plocal} -- displayed on a color log scale as a function of lattice amplitude $A$ and superlattice amplitude $B$. 
  The delocalization is less than one percent above the solid line. The dashed line is the same as that in Fig.~\ref{fig:plz} and 
  shows where the Landau-Zener crossing has an error probability of less than 
  one percent. The hatched area marks parameters where the dynamics cannot be well described by Landau-Zener physics.}
  \label{fig:plocal}
\end{figure}

The results are shown in Fig.~\ref{fig:plocal}. As seen, the delocalization error decreases with increasing superlattice potential strength, $B$ because a stronger superlattice reduces the coupling to the neighboring sites.
The delocalization error is less than 1\% above the solid line. 
We  also indicate by a hashed line the region where the Landau-Zener error is less than 1\% as also shown in  Fig.~\ref{fig:plz}. In this way, the solid and hashed lines 
define a region in $(A,B)$ space which is well suited for the SCB procedure, since both the delocalization error and the Landau-Zener error is less than 1\%. 
 
\subsection{Decreasing errors between neighbor well degeneracies -- the modified SCB} \label{sub:IID}
The two previous sections analyzed the possible errors associated with the traversal of the  degeneracy point, and determined a range of values for the lattice parameters $(A,B)$, where the SCB procedure works reasonably well. 
Now, we discuss the dynamics when the superlattice is in-between degeneracy points,  as illustrated in Fig.~\ref{fig:procedure}(b). We will first show that certain complications arise during this step in the original SCP procedure and then discuss how the procedure can be modified to yield an almost perfectly working conveyor belt by turning down the superlattice potential, when it is in-between the degeneracy points.

\begin{figure}[tb]
  \centering
  \includegraphics[width=\columnwidth]{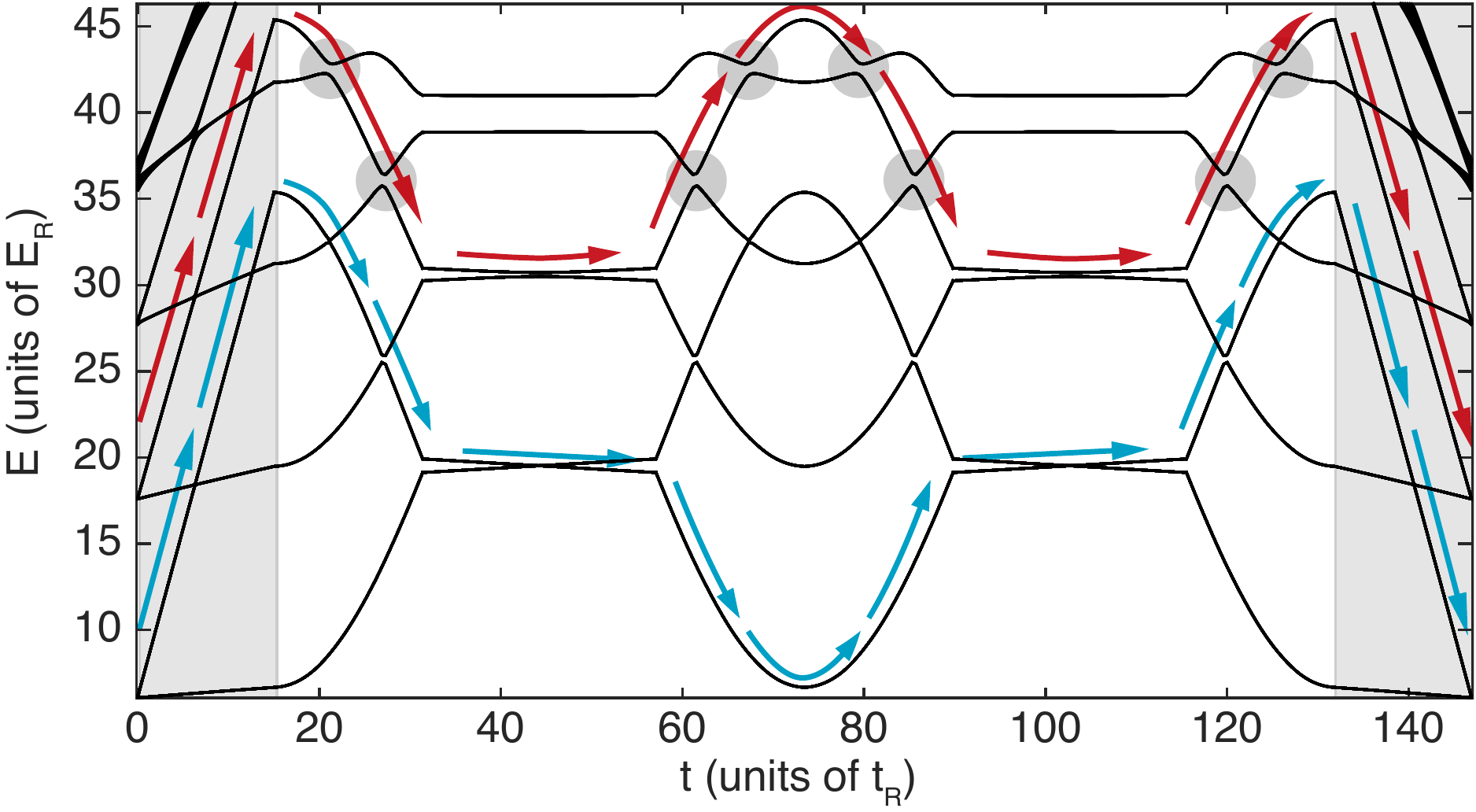}
  \caption{Eigenenergies of the original superlattice procedure with $A = 40 E_R$ and  $B = 30 E_R$. Arrows show the desired path taken by the ground state and first excited state starting from the even lattice sites. The filled grey circles mark unwanted resonances between the ground state, first excited state, and higher excited vibrational states.}
  \label{fig:SCB-A40B30}
\end{figure}

As seen from Fig.~\ref{fig:plocal}, the superlattice amplitude $B$ must be comparable to the lattice potential, $A$ in order to avoid the delocalization of the first excited state during the traversal of a neighbor well degeneracy. A large superlattice potential, however, causes errors when moving between consecutive neighbor well degeneracies. 
During this step, we want all the atoms to remain at the same lattice position but delocalization errors as well as unwanted resonances with higher excited states in the neighbor well can diffuse the atomic wave functions.  Examples of such unwanted resonances are shown in Fig.~\ref{fig:SCB-A40B30}, where the undesired resonances are marked with grey circles. These errors are particularly large during the displacement if a strong superlattice potential is used since the superlattice pushes one of the states towards being almost unbound as shown in the right hand side of Fig.~\ref{fig:procedure} (b). This can directly be seen by comparing Fig.~\ref{fig:SCB-A40B20} with Fig.~\ref{fig:SCB-A40B30}. The latter has a higher value of $B$ and hence more and stronger avoided crossings. 

The unwanted effects discussed above can be decreased by speeding up the superlattice potential, ensuring that the wave function has no time to delocalize and that any avoided crossings, associated with the unwanted resonances, are traversed diabatically~\cite{Pedersen2011}. 
The superlattice velocity cannot, however, be increased indefinitely, since this induces uncontrolled excitations to higher Bloch bands if the states cannot follow the bands. Hence one must find optimal crossing velocities which balances in between both sources of errors. For parameters such as the ones used in Fig.~\ref{fig:SCB-A40B30} the avoided crossing are so strong that we have not been able to find a suitable velocity where the SCB has a good performance. For the original proposal of Ref. \cite{Pedersen2011} we have therefore only been able to identify  a small region of parameter space around ${(A,B)= (40 E_R,20 E_R)}$ where the procedure works  well. Even  for these parameters, however, 
the error is considerable. For atoms in the first excited state occupying the sub-lattice site which is lifted in energy by the superlattice [the red atom on the right hand side of Fig.~\ref{fig:procedure} (b)] the error grows to approximately 60\% after 7-8 repetitions~\cite{Pedersen2011}.

Fortunately, the source of error discussed above can be almost eliminated by using a modified superlattice ``conveyor belt'' (MSCB) method, which only uses a strong superlattice potential in the vicinity of each degeneracy point. By ramping down the superlattice in between the degeneracy points there is strongly reduced  tunneling of the atoms to unwanted sites. The ramped down superlattice potential  is shown in Fig.~\ref{fig:procedure}(d), which is the MSCB replacement of  Fig.~\ref{fig:procedure}(b) of the original SCB. 

\begin{figure}[tb]
  \begin{center}
  \includegraphics[width=0.5\textwidth]{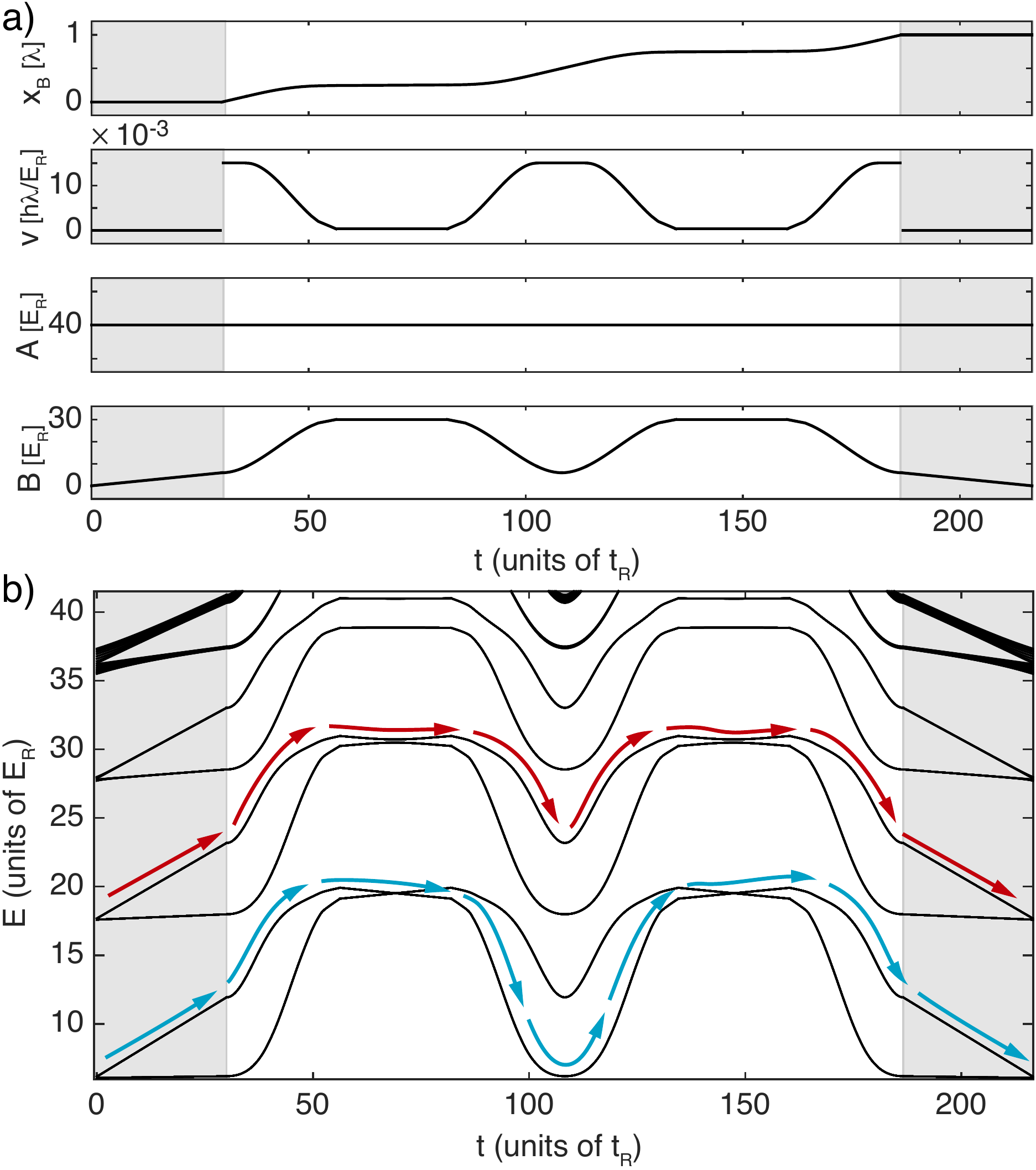}
  \caption{The modified superlattice conveyor belt procedure 
  with neighbor well degeneracies traversed at $A = 40 E_R$ and $B= 30 E_R$. (a) The lattice-superlattice parameters  as a function of time. (b) The instantaneous Bloch 
  energies as a function of time. Note how the intermediary ramp down of the superlattice potential, $B$, removes unwanted energy 
  crossings present in the original procedure of Fig.~\ref{fig:SCB-A40B30}.}
  \label{fig:MSCB-A40B30}
  \end{center}
\end{figure}

As a tangible example, Fig.~\ref{fig:MSCB-A40B30}(a) shows the time-evolution of the superlattice position, $x_B$, the superlattice velocity, $v_B$, and the superlattice intensity for the modified procedure when neighbor well degeneracies are crossed at lattice-superlattice amplitudes ${(A,B)=(40 E_R, 30 E_R)}$. 
Both the superlattice amplitude and the superlattice velocity is changed smoothly in order to decrease unwanted  excitations. Figure~\ref{fig:MSCB-A40B30}(b) shows the instantaneous eigen-energies of the superlattice Bloch waves as a function of time for the MSCB. In contrast to the original SCB procedure of Fig.~\ref{fig:SCB-A40B20}, the unwanted crossings of the Bloch bands have now been removed by the dynamical ramp down of the superlattice amplitude. 
This eliminates  the main sources of errors in-between degeneracy points, and the accuracy of the MSCB method is now mostly limited by the physics of the adiabatic/diabatic evolution near the degeneracy point.

\section{Numerical calculations} \label{sec:III}
In order to quantitatively test our predictions concerning the MSCB, we numerically calculate the single-particle dynamics of the system by solving the time-dependent Schr\"{o}dinger equation. As explained above, we assume that the  atomic wave functions remain in the vibrational ground state in the perpendicular $(y,z)$-directions. 
This assumption is valid if the total time taken to perform the superlattice procedure is much smaller than the wave function dephasing time along these directions, which can be realized experimentally by using a tight lattice potential in the $(y,z)$-directions. Working in superlattice units (i.e. measuring x in terms of $\lambda$, energies in terms of $E_R$ and times in units of $t_R$), the one dimensional Schrödinger equation describing the system is then 
\begin{align}
  [ - \frac{1}{4 \pi^2} \partial^2_x +V_A(x)+ V_B(x,t)] \psi(x,t)
  =
  \frac{i}{2 \pi} \partial_t \psi(x,t).
\end{align}

We discretize the Schr\"{o}dinger equation on a real-space grid with 128 points and solve the dynamics by expanding a a basis consisting of the  16 lowest lying eigenstates of the underlying lattice $V_A$.  
The MSCB procedure is pieced together from basic steps, $1\ldots i \ldots m$, defined over a range of superlattice positions $x_B \in [x_B(i), x_B(i+1)]$. The steps fall into three categories: the initial and final superlattice ramps, the dynamics close to the degeneracy point, and the dynamics in-between degeneracy points. 
The dynamics is  calculated as evolution matrices $U$ for each of these basic steps of the MSCB procedure,  as discussed in Appendix~\ref{app:U}. 
This allows for an efficient calculation of the total evolution simply by multiplying the evolution matrices for each individual step together. 
In the original SCB, the superlattice velocity and superlattice intensity were kept constant within each  step, but as illustrated in Fig.~\ref{fig:SCB-A40B20}, they were allowed to change abruptly between consecutive steps. In contrast, the MSCB only allows the superlattice parameters to vary smoothly in time. Specifically, we chose the superlattice parameters to be third order polynomials in time with vertical tangents at the end points of each basic step. As explained in Appendix~\ref{app:param} this amounts to a total of 9 parameters including the optimum superlattice velocity, which can be changed in order to find the optimal MSCB procedure. 

We choose the lattice-superlattice amplitudes at the degeneracy point, and use the optimum superlattice velocity calculated from Eqs.~\eqref{eqn:plz}-\eqref{eqn:PNWD} as a reasonable starting point. The remaining parameters are chosen so as to maximize the total localization probability of the atoms in the first excited states, estimated as
\begin{align}
  P_{\text{local}}^l = \left|\frac{1}{N} \sum_{k} e^{- i \int_0^t E_k^l (t') \mathrm{d}t'} \right|^2.
  \label{eqn:Pdelocal}
\end{align} 
In order to measure the quality of the MSCB we define the error probability as the amount of a wave function which does not end up in the desired place. The ground state wave functions, $l=1$, should remain put, while the first excited state wave functions, $l=2$, should be moved a number of lattices sites, $m$, corresponding to the number of traversed degeneracy points, $m$. 

Fixing the lattice-superlattice amplitudes $(A,B)$ at the degeneracy point, we perform multiple simulations of the MSCB each time varying one or more of the remaining parameters. Because of its vastness we cannot map out the entire parameter space, and we are satisfied when we find a local minimum. 

\begin{figure}[tb]
	\begin{center}
	\includegraphics[width=\columnwidth]{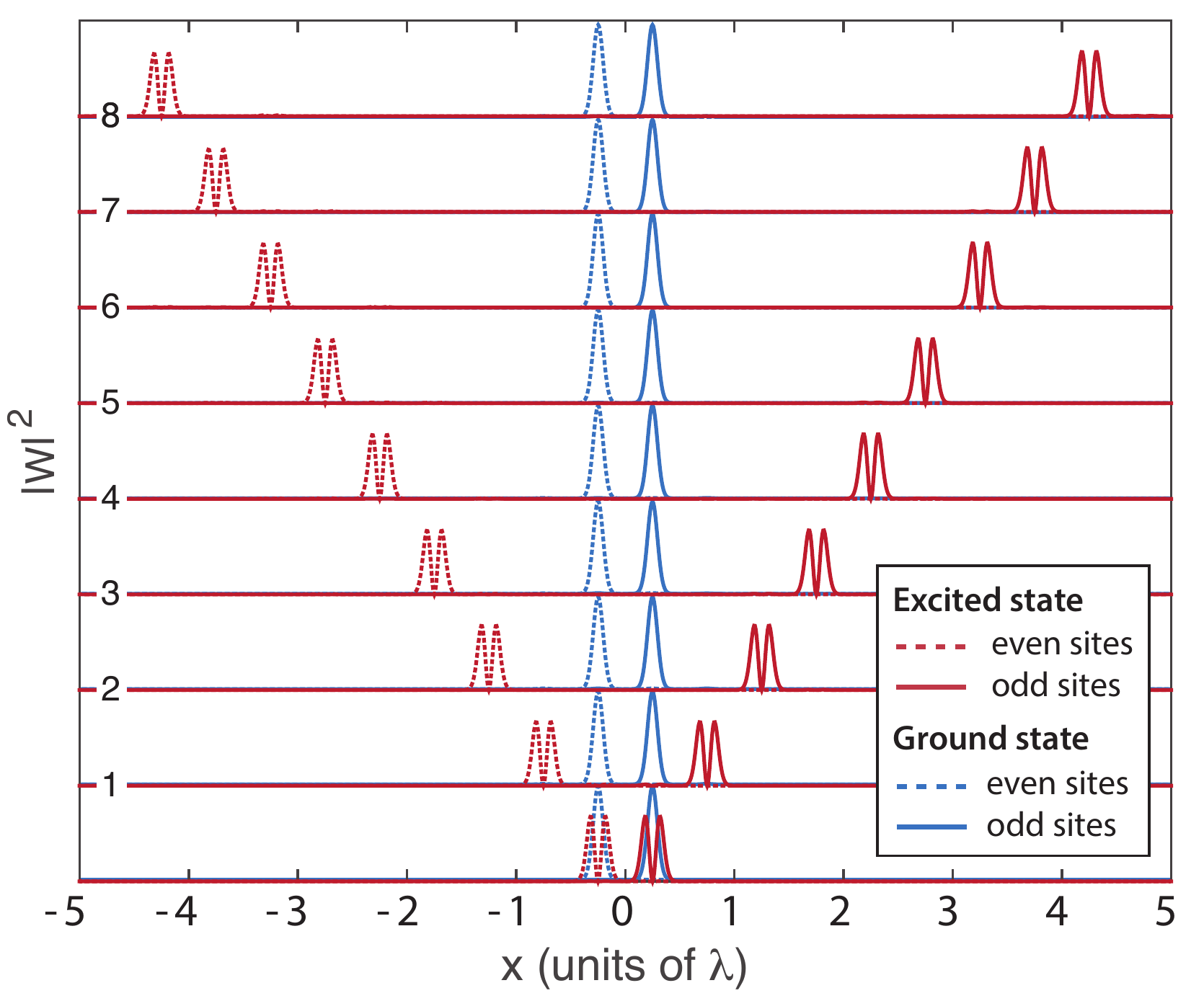}
	\end{center}
	\caption{The ground state and excited state Wannier functions of even and odd lattice states after 1-8 repetitions of the MSCB procedure where the degeneracy point is traversed with lattice-superlattice amplitudes $(A,B) = (40 E_R, 30 E_R)$.}
	\label{fig:wannier}
\end{figure}

In this way, we find that the MSCB procedure works best for degeneracy point lattice strengths ${(A, B) = (40 E_R, 30 E_R)}$, where it is essentially without error. 
This is illustrated  in Fig.~\ref{fig:wannier}, where we plot the wave functions after up to eight lattice translations with $(A, B) = (40 E_R, 30 E_R)$.
We see that the vibrational ground state wave function is essentially unperturbed, whereas the first vibrational states are translated up to eight lattice positions as desired. 
\begin{figure*}[tb]
	\begin{center}
	\includegraphics[width=\textwidth]{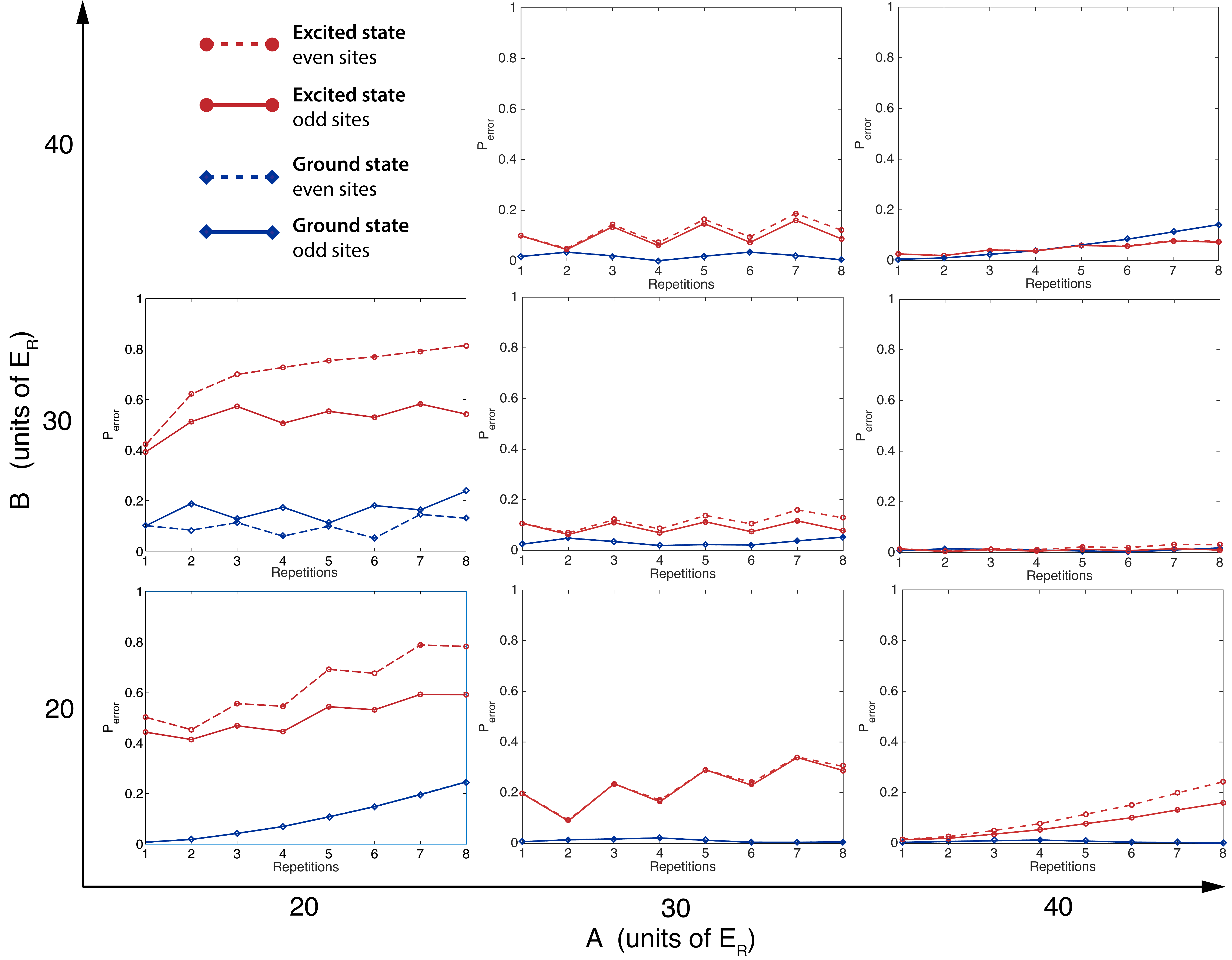}
	\end{center}
	\caption{The error probability, given as the wave-function weight left outside the desired final state, for the MSCB procedure for various lattice amplitudes $A$, and superlattice amplitudes $B$. Each 
	plot shows the error-probability as a function of lattice translations for atoms in the ground vibrational state and in the first excited  state. Note that the MSCB procedure distinguishes between atoms on even and odd lattices sites as determined by the initial position of the superlattice. The lattice parameters with the smallest error are
	$(A, B) = (40 E_R, 30 E_R)$, for which  the central crossings can be optimized almost perfectly, and the disturbances in-between these crossings can be almost eliminated. The parameterizations of each of these MSCB scenarios are explicitly stated in Appendix~\ref{app:param}.}
	\label{fig:psuccess}
\end{figure*}

The error probability of each of the four relevant initial states obtained for various amplitudes of the lattice and the superlattice is plotted in Fig.~\ref{fig:psuccess}. 
The excellent performance of the MSCB procedure with ${(A, B) = (40 E_R, 30 E_R)}$ is  evident. In addition, many of the other scenarios perform as well as the best example of the original SCB procedure described in~\cite{Pedersen2011}. As can be seen, the simulated errors diminish with increasing values of $A$ or $B$, which is in qualitative agreement with the error estimates of Fig.~\ref{fig:plz} and Fig.~\ref{fig:plocal}. Only the extreme case ${(A, B) = (40 E_R, 40 E_R)}$ abhors from this picture, because the large superlattice potential makes it impossible to separate the different levels when moving between degeneracy points.

\section{Conclusion and discussion}
We have carefully analyzed the different steps in the superlattice conveyor belt and identified the limiting factors and potential sources of errors. 
Based on this, we have been able to optimize  and improve the performance  by dynamically tuning the superlattice potential during the procedure. The main features of our findings obtained by simple model calculations, were confirmed by a direct simulation of the system dynamics for various lattice-superlattice parameters. Based on the optimization, we found an almost ideal performance of the conveyor belt for lattice parameters around $(A, B) = (40 E_R, 30 E_R)$. As a result the probe can be used to perform an essentially error free measurement of magnetic correlations in only a single shot of the experiment by measuring the fraction of doubly occupied sites. As opposed to experiments based, e.g., on measuring  noise correlations~\cite{Altman2004,Bruun2008,Bruun2009} where an average has to be performed over multiple shots, this is a major advantage of this proposal. We thus believe that the procedure can be a valuable tool for future optical lattice experiments. 
 
From an experimental perspective one of the main challenges of the proposed method is that it  requires a precise phase lock between the lattice laser potential and the superlattice laser potential. Under the ideal conditions $(A, B) = (40 E_R, 30 E_R)$  we have assumed that the cross over region in the adiabatic passage has a width of $w\sim 0.009 \lambda$. Hence the relative position of the two lattices needs to be controlled with great precision. Possibly this requirement can be reduced by working with higher excited vibrational states or by extending the region of the slow dynamics. The latter will, however, come at the prize of a longer operation time of the procedure. Another challenge for the proposed method arises from the fact that in most experimental systems the lattice height is not homogeneous but varies across the sample.  Hence the lattice parameters cannot be kept at ideal working conditions for all positions in the lattice. Here it is a major advantage of our modified scheme that it is much more robust to variations than the original scheme which only worked in a small region in parameter space. In the end the procedure is based on Landau-Zener transitions which are not very sensitive to the precise value of the tunneling.  As long as the parameters do not show too strong variations, we therefore believe that it should be possible to apply the modified procedure even for systems which are not completely homogeneous. 

\section{Acknowledgements}

The research leading to these results has received funding from the European Research Council under the European Union’s Seventh Framework Programme (FP/2007- 2013)/ERC Grant Agreement No. 306576. B. M. A. acknowledges financial support from a Lundbeckfond fellowship (Grant No. A9318). G.M.B. wishes to acknowledge the support of the Villum Foundation via grant VKR023163.   

\appendix

\section{Parametrization of the MSCB} \label{app:param}
\begin{figure}[tb]
	\centering
	\includegraphics[width=\columnwidth]{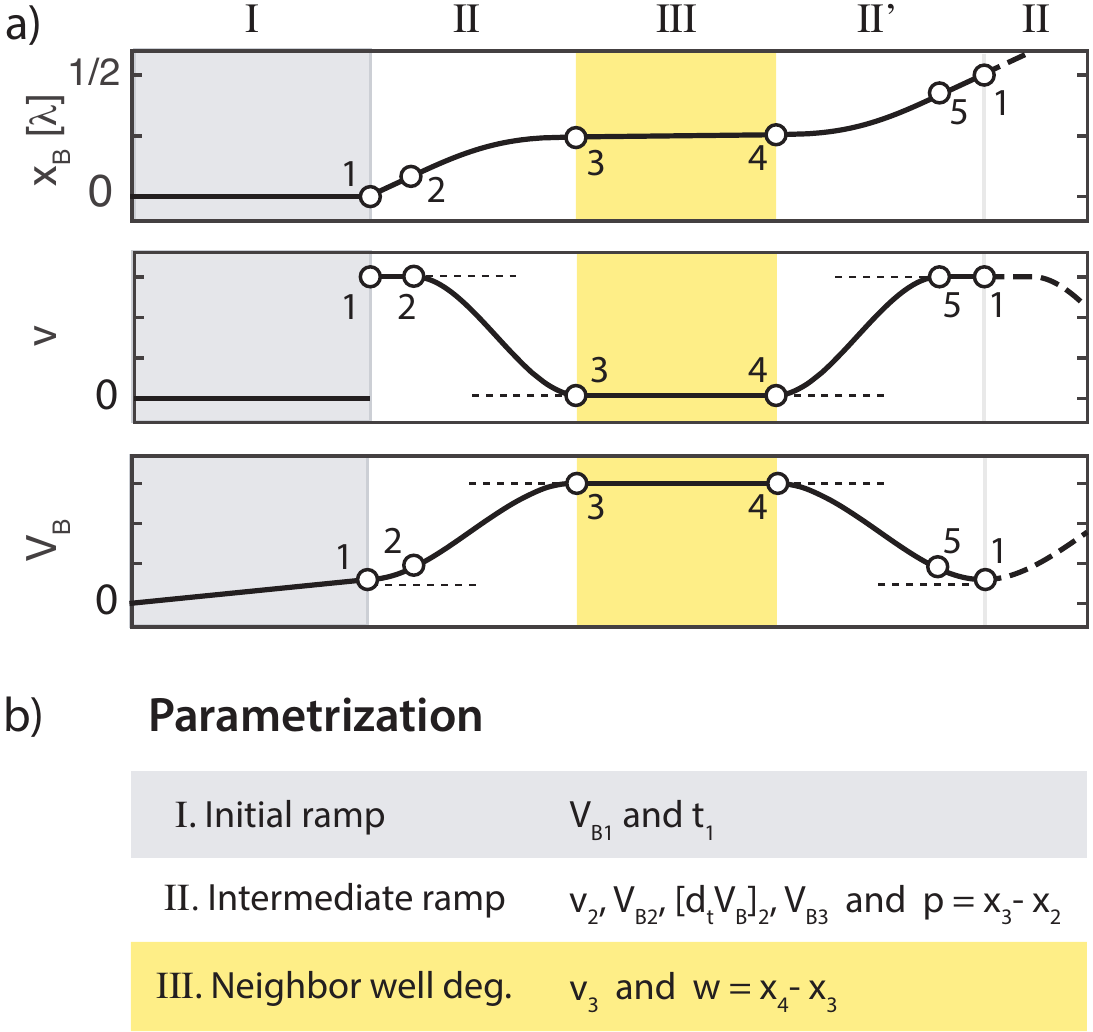}
	\caption{ a) The parametrization of the modified SCB procedure. The five control points for the three different stages of the MSCB: the initial ramp, the intermediate ramp and the neighbor well degeneracy. Note that the procedure is symmetric around the degeneracy points. Circles mark the five control points and dashed lines indicate the horizontal tangents at the end points of the third order polynomial curves. b) Summary of the parametrization of the MSCB for each of the three stages.}
	\label{fig:procedural}
\end{figure}

Here we describe the detailed parameterization of the MSCB, which was used to obtain the results in Figs.~\ref{fig:wannier} and \ref{fig:psuccess}. The parameters and basic steps are illustrated in Fig.~\ref{fig:procedural}. 
We parameterize the modified SCB procedure in terms of five different control points. The first point controls the initial step I with a slow ramp up of the superlattice, where the intensity is increased to some small value $B_1$ over a period $t_1$. The initial ramp is followed by step II, where the superlattice is further ramped up while being simultaneously moved towards the next degeneracy point. The superlattice position is advanced at a high velocity, $v_2$, until it reaches the second control point, after which the velocity is smoothly ramped down in order to the match the optimal velocity, $v_{3}$, upon the arrival at the third control point. Specifically we model the velocity ramp as a third order polynomial in time with horizontal tangents at the initial and the final control points. At the same time the superlattice intensity is increased further. This is done smoothly by following a third order polynomial in time with a horizontal tangent at the initial control point. 

At the second control point, we choose a superlattice intensity $B_2$ and a slope $[dB/dt]_2$. The ramp continues smoothly in time from the second to the third control point and ends at a superlattice intensity, $B_3$ - again with a horizontal tangent. We chose step II to happen over a range of superlattice positions, $p = x_3-x_2$ (see Fig.~\ref{fig:procedural}), during a total time, $T = 2 p/(v_3 - v_1)$. 

At the next step III, the  degeneracy point is traversed at a constant superlattice intensity, $B_3$, and velocity, $v_3$. After traversing the degeneracy point we have another step II', where we decrease the superlattice intensity and increase the velocity. This is done in a completely similar fashion as the ramp before the degeneracy point in step II by constructing the time reverse of that procedure with the same parameter values. 
This ensures that the procedure is symmetric with respect to the degeneracy points at superlattice positions $x_B = \lambda/4 + n \lambda/2$, where $n$ is an integer.

As described in the main text, we choose fixed values of the lattice-superlattice amplitudes $(A,B)$ at the degeneracy point. We then estimate the optimal Landau-Zener velocity by optimizing Eq.~\ref{eqn:PNWD}. The remaining parameters are initially chosen such as to minimize the overall delocalization error as estimated by Eq.~\ref{eqn:Pdelocal}. From this initial point in the impossibly large parameter space, we perform multiple simulations where we vary each parameter slightly in order to find a local minimum of the total error probability. 

In Figure~\ref{fig:psuccess} we showed the error probability of the best found parameterization of the SCB for different lattice-superlattice amplitudes $(A,B)$ at the degeneracy point. We supply the parameterizations for each $(A,B)$ pair in the table below:

\vspace{.2cm}
\noindent
{\small \begin{tabular}{|c||c|c|c|c|c|c|c|c|}
\hline
$(A,B_3)$ &  $V_{B1}$ &  $t_1$ &  $v_2$ &  $V_{B2}$ &  $\mathrm{d}_t V_B|_2$ &  $p$ &  $w$ &  $10^3 v_3$ \\
$[E_R]$   &  $[E_R]$ &  $[t_R]$ &  $[\lambda/t_R]$ & $[E_R]$ & $[E_R/t_R]$& $[\lambda]$ & $[\lambda]$ & $[\lambda/t_R]$ \\
\hline
(20, 20) &   2.0 &      2.0 &  0.5   &  6.00  &                  35.3 & 0.1  & 0.03  & 150      \\
(20, 30) &   2.0 &      0.4 &  0.5   &  6.00  &                  54.3 & 0.1  & 0.10   & 150      \\
(30, 20) &   0.5 &      1.0 &  0.1   &  6.00  &                  9.55 & 0.1  & 0.01  & 2.0    \\
(30, 30) &   6.0 &      6.0 &  0.15  &  6.47  &                  3.92 & 0.2  & 0.03  & 4.5  \\
(30, 40) &   1.0 &      1.0 &  0.05  &  6.00  &                  5.85 & 0.08 & 0.035 & 40.0  \\
(40, 20) &   1.0 &     10.0 &  0.005 &  9.00  &                  0.95 & 0.08 & 0.004 & 0.12 \\
(40, 30) &   6.0 &     30.0 &  0.015 &  8.90  &                  1.00 & 0.16 & 0.009 & 0.35 \\
(40, 40) &   8.0 &      8.0 &  0.1   &  8.95  &                  1.00 & 0.2  & 0.012 & 0.30\\
\hline
\end{tabular}}

\section{Evolution matrices} \label{app:U}
In this appendix we describe the calculation of evolution matrices, which forms the basis of our numerical simulation. Consider the matrix solutions ${U}(t,0)$ to the ordinary partial differential equation
\begin{equation} \label{floquets_pde}
  \dot{{U}} = A(t) U,
\end{equation}
where $A(t) = A(t+T)$ is a periodic time-dependent matrix with a period $T$. The equation $\dot{x} = A (t) x$ with the boundary condition $\vec{x}(0) = \vec{x}_0$ is solved directly by $\vec{x}(t) = U(t,0) \vec{x}_0$ if we impose the initial condition $U(0,0)=1$. Applying the periodicity of $A$ we see that the $U$ matrix can be decomposed into
\begin{equation}
  U(t+mT,0) = U(t,0) U(T,0)^m, \qquad m \in \mathbb{N}.
  \label{eqn:floquets}
\end{equation}
In our simulation we solve the evolution within each Bloch wave sector separately, from
\begin{align}
\dot{\Phi}^k_{\mu} (t) = - i H(t) \Phi^k_{\mu} (t),
\end{align}
where $k$ is the Bloch wave number, and $\mu$ is a level index. Because the Bloch bands do not mix, we find the general evolution matrix elements between different levels, $\mu$ and $\nu$, for each wave number $k$,
\begin{equation}
  U_{\mu \nu}^k (t_1, t_0) = \langle \Phi_{\mu}^k (t_1) | \Phi_{\nu}^k (t_0)\rangle.
\end{equation}
In some common basis, $|\tilde{\Phi}_{\eta}^k \rangle$, we have
\begin{equation*}
  \tilde{U}_{m n}^k (t_1, t_0) = \sum_{\mu \nu} \langle \tilde{\Phi}_{m}^k | \Phi_{\mu}^k (t_1) \rangle  U_{\mu \nu}^k (t_2, t_1) \langle \Phi_{\nu}^k (t_1) | \tilde{\Phi}_{n}^k \rangle,
\end{equation*}  
and they are naturally concatenated by matrix multiplication in order to yield the transformation matrix for the compound evolution,
\begin{equation}
  \tilde{U}_{ln} (t_2, t_0) = \tilde{U}_{lm} (t_2,t_1) \tilde{U}_{m n} (t_1,t_0).
\end{equation}
Each step of the procedure is then solved independently, multiplied together and extended to multiple periods through the use of Eq.~\ref{eqn:floquets}. The final evolution can be transformed to a real-space position basis by rewriting $\tilde{U}_{mn}^k$ in a basis of Wannier functions,
\begin{align}
  |\Psi_{m}^R\rangle
  &=
  \frac{1}{\sqrt{N_k}} \sum_k e^{i k R} |\tilde{\Phi}^k_{m}\rangle,
\end{align}
such that the real-space evolution matrix becomes,
\begin{equation}
  \tilde{U}^{RR'}_{m n}(t,0) = \frac{1}{N_x} \sum_k e^{ik(R-R')} \tilde{U}^k_{mn} (t,0).
\end{equation}

Choosing the common basis, $|\tilde{\Phi}_{m}^k \rangle$, to coincide with the initial (and final) eigenstates of the system, we can directly identify the probability that an atom in some initial Wannier state end up another Wannier state after the time evolution,
\begin{align}
  \langle \Psi^{R'}_{\mu} (t) | \Psi^{R_0}_{\eta} (t_0) \rangle
  &=
  \tilde{U}^{R R_0}_{\mu \eta}(t, t_0).
\end{align}
The success probabilities can be calculated as the norm square of the appropriate matrix elements. Starting in the ground-state $n = 0$ at a position $R_0=8$, this atom should remain put and the success-probability is given by ${P_\text{success} = |U^{88}_{00}(t)|^2}$. When instead starting in the excited state $\mu_0=1$, the atom should move along the lattice, and the success-probability after an even number of repetitions, $m$, is given by $P_\text{success} = |U^{8-m/2,8}_{11}|^2$. 

\section{Correlation functions} \label{app:II}
The correlation functions measured by the SCB can be determined in the following way. At an initial time $t_0$ we begin the application of the SCB, which ends at a later time, $t_1$. The operator $\create{a}{\sigma i} (t)$ creates a particle with spin $\sigma$ at site $i$ at time $t$. The workings of the SCB is then described by transition matrix elements, defined by
\begin{align}
  \create{a}{\sigma i} (t_1) &= \sum_j U_{\sigma i j}^* \create{a}{\sigma j} (t_0).
\end{align}
Write ${a = a(t_0)}$. The probability that the site $i$ is doubly occupied after the sweep so a molecule can be formed is $P_{mol} = \langle \n{i \uparrow} (t_1) \n{i \downarrow} (t_1) \rangle$, where $\n{i} = \create{a}{i} \annihilate{a}{i}$. In the Mott phase we have that
\begin{align}
  P_{mol} &
    = \sum_{jl} U_{\uparrow ij}^{*} U_{\uparrow ij} U_{\downarrow il}^{*}  U_{\downarrow il}
      \langle \n{\uparrow j} \n{\downarrow l} \rangle
      \nonumber \\ &\quad 
      \pm 
      \sum_{j \neq l} U_{\uparrow ij}^* U_{\uparrow il} U_{\downarrow il}^* U_{\downarrow ij} \langle \create{a}{\uparrow j} \annihilate{a}{\downarrow j} \create{a}{\downarrow l} \annihilate{a}{\uparrow l} \rangle.
\end{align}
Introduce the spin operators $\n{\uparrow} = 1/2 + s^z$, $\n{\downarrow} = 1/2 - s^z$, $s^+ = \create{a}{\uparrow} \annihilate{a}{\downarrow}$, and $s^- = \create{a}{\downarrow} \annihilate{a}{\uparrow}$. Then
\begin{align}
  P_{mol} &
    = \sum_{jl} U_{\uparrow ij}^{*} U_{\uparrow ij} U_{\downarrow il}^{*}  U_{\downarrow il}  (1/4 - \langle s^{z}_{j} s^z_l \rangle) 
      \nonumber \\ &\quad 
      \pm
      \sum_{j \neq l} U_{\uparrow ij}^* U_{\uparrow il} U_{\downarrow il}^* U_{\downarrow ij} \langle s^{+}_j s^-_l \rangle.
\end{align}
Here we assumed $\langle s^z \rangle =0$. Knowing the transition matrices one can easily calculate the double occupancy if knowing $\langle s^z_{j} s^z_{l} \rangle$ and $\langle s^{+}_j s^-_l \rangle$ from e.g. Quantum Monte Carlo simulations.

When the spin down species is pinned we have ${U_{\downarrow ij} = \delta_{ij}}$, and 
\begin{align}
  P_{mol} &=
    \sum_{j} U_{\uparrow ij}^{*} U_{\uparrow ij} (1/4 - \langle s^z_j s^z_i \rangle).
\end{align}
If the spin up species is transferred perfectly $m$ sites along the lattice, $U_{\uparrow ij} = \delta_{i,i+m}$, and
\begin{align}
	P_{mol} &= 1/4 - \langle s^z_{i+m} s^z_{i} \rangle.
\end{align}

\bibliography{refs}

\end{document}